# The Combinational Mutation Strategy of Differential Evolution Algorithm for Pricing Vanilla Options and Its Implementation on Data during Covid-19 Pandemic

Werry Febrianti 1,2,*, Kuntjoro Adji Sidarto 1,* and Novriana Sumarti 1,*

1 Department of Mathematics, Faculty of Mathematics and Natural Sciences, Institut Teknologi Bandung, Ganesa Street No. 10, Bandung, 40132, Indonesia; werry_febrianti@students.itb.ac.id, sidarto@itb.ac.id, novriana@itb.ac.id

2 Mathematics, Department of Sciences, Institut Teknologi Sumatera, Terusan Ryacudu Street, Way Huwi, South Lampung 35365, Indonesia; werry.febrianti@ma.itera.ac.id

* Correspondence: werry.febrianti@gmail.com (W.F.), sidarto@itb.ac.id (K.A.S.), novriana@itb.ac.id (N.S.)

**Abstract**

Investors always want to know about the profit and the risk that they will be get before buying some assets. Our main focus is getting the profit and the probability of getting that profit using the differential evolution algorithm for vanilla option pricing on data before and during COVID-19 pandemic. Therefore, we model the pricing of an option using a bi-objective optimization problem using data before and during COVID-19 pandemic for one year expiration date. We change this problem into an optimization problem using adaptive weighted sum method. We use metaheuristics algorithm like Differential Evolution (DE) algorithm to solve this bi-objective optimization problems. In this paper, we also use modification of Differential Evolution for getting Pareto optimal solutions on vanilla option pricing for all contract. The algorithm is called Combinational Mutation Strategy of Differential Evolution (CmDE) algorithm. The results of our algorithm are satisfactory close to the real option price in the market data. Besides that, we also compare our result with the Black-Scholes results for validation. The results show that our results can approximate the real market options more accurate than Black-Scholes results. Hence, our bi-objective optimization using Combinational Mutation Strategy of Differential Evolution algorithm can be used to approximate the market real vanilla option pricing before and during COVID-19 pandemic.

## 1. Introduction

Financial derivative problems are usually complex problems, dynamics, and uncertain. One of interest in derivative market is determining the benefit of an option pricing. Based on the price, maturity time, rate of the return, and risk of neutral interest rate, an investor can predict the value of the real market data before buying/selling the option value contract.

Metaheuristics algorithms were used to determine option pricing value (Kumar, et al. 2014). One of metaheuristics algorithms that we have chosen to use in this paper is Differential Evolution (DE) algorithm. DE can be a robust tool to find the minimum solution for these option value problems. Besides that, DE algorithm is easy in implementation and very efficient in exploration and exploitation strategies. In this paper, we also use modification of Differential Evolution Algorithm which is called

Combinational Mutation Strategy of Differential Evolution Algorithm (CmDE) to find the solution of this bi-objective optimization problems. CmDE is a modification of DE by Febrianti et al. 2022. CmDE has been used for solving stiff ordinary differential equations. Febrianti et al. 2022 can approximate the solution of stiff ordinary differential equations using optimization method with CmDE algorithm. The results show that CmDE can be a good tool for solving stiff ordinary differential equations.

Black, Scholes, and Merton (BSM) introduce an option pricing calculation (Black, & Scholes, 1973; Merton, 1973). BSM presented an analytical solution in pricing simple derivative with simplifying assumption such as constant volatility, but this BSM model did not appropriate for the other options. Therefore, there is an expansion to exercise the option if the maturity time is not in the end of the contract but in the middle of the contract. This option is called an American option which uses binomial model to solve it. Binomial model presents the calculation for American option by using the same assumption with the BSM model. There are many literatures and practical evaluation of derivative instruments which enriched many quantitative techniques (Hull, 2014; Jiang, & Li, 2005; Brennan, & Schwartz, 1978; Cerny, 2004). But, these techniques can not able to find the appropriate solution that near with the market.

Based on accuracy and efficiency to approximate option pricing in approaching option pricing, we propose a technique for modelling an option pricing into an optimization problem (Deb, 2001; Deb, et al. 2002; Singh, et al. 2016). Thus, we use DE for finding the solutions.

The next section describes the concept of financial option. In Section 3, DE algorithm is explained. In Section 4, the explanation of the strategy of our proposed about using the DE algorithm with adaptive weighted sum method to vanilla option pricing. In Section 5, we describe about the way that we used to get the data and the results of the implementation of this algorithm are described along with evaluation and error analysis using data before the COVID-19 happened and when COVID-19 happened. The last section is about conclusion.

## 2. Option Pricing

Option pricing is a contract between two persons: writer and holder until expiry date. Writer is a person who sells the contract of the option. Holder is a person who buys the contract of the option. Based on the right to own the asset, options are divided into two types: call and put. A call option is an option that gives right to the holder for buying that underlying asset in maturity time. Then, a put option is an option that gives right to the holder to sell that underlying asset in maturity time. Then, based on maturity time and the way of the option to be exercised, option consists of vanilla options (European, American) and exotic options (Asian, Barrier, Lookback, etc.). At this time, we only focus on solving vanilla call option pricing.

Vanilla option pricing has a purpose to find the value of European and American call options. The underlying stock price ($S$), strike price ($K$), interest rate ($r$), volatility ($\sigma$), and maturity time ($T$) are parameters that is involved in calculating this vanilla option pricing. We use max ($S - K$, 0) and max ($K - S$, 0) for calculating call and put options, respectively.

**Algorithm 1.** CmDE algorithm

---
1 : Determining of parameters for DE.
2 : **Initialization** Generate the initial population $X_{i,G} = \{x_{i,G}^1, x_{i,G}^2, \ldots, x_{i,G}^D\}, i = 1, 2, \ldots, NP$
3 : Assess the fitness for each individual
4 : **while** Termination condition is not satisfied **do**
5 :   **Mutation** set $\lambda = 0.5$ and calculate each mutation (1)-(3)
6 :   **Crossover**
7 :   Evaluate the boundary constraints for each $v_{i,G+1}^j \in X_{i,G+1}$ in each mutation strategy
8 :   **Particular Selection**
     Select the best population of each mutation strategy
9 : **end while**
10: **Global Selection**
    Select the best solution of all mutation strategy
---

## 3. Differential Evolution Algorithm

Differential Evolution (DE) is one of the global optimization methods that was first introduced in 1995 by Storn and Price (Price, Storn, & Lampinen, 2005). DE has various schemes describe as DE/*a*/*b*/*c* where *a* is the mutation vector, *b* is total of different pair vectors in the mutation, and *c* is crossover's type. Some detail explanation on these schemes can be found in (Febrianti, Sidarto, & Sumarti, 2021).

Mutation, crossover and selection are the steps in DE. Then, the "DE/rand/1/bin" classical mutation strategy in a $G$ generation of populations has formula:

$$v_{i,G+1} = x_{r_1,G} + \lambda(x_{r_2,G} - x_{r_3,G}) \tag{1}$$

There are three different vectors, where $x_{r_1,G}$ is a base vector, and vectors $x_{r_2,G}, x_{r_3,G}$ are used for its difference. All vectors are randomly chosen from the populations, stated by "rand". The scale factor $\lambda$ is a scale factor in mutation that is a constant between 0 and 1. "bin" stands for binomial crossover, and "1" stands for a number of pair of different vectors in Equation (1). Based on our experience, when the base vector ($x_{r_1,G}$) is changed into ($x_{best,G}$), the results become better in finding the optimum solutions. Therefore, in our mutation scheme in DE algorithm is using combinational mutation strategy, so our algorithm called CmDE. Then, in our mutation algorithm, the scale factor $\lambda$ is chosen to be 0.5 and the strategy of our mutations is like (1) and below:

DE/best/1/bin: $v_{i,G+1} = x_{\text{best},G} + \lambda(x_{r_1,G} - x_{r_2,G})$ (2)

DE/current-to-best/1/bin: $v_{i,G+1} = x_{i,G} + \lambda(x_{\text{best},G} - x_{i,G}) + \lambda(x_{r_1,G} - x_{r_2,G})$ (3)

The CmDE algorithm is stated in Algorithm 1.

A vector's target ($x_{i,G}$) can be potentially directed by a mutant vector ($v_{i,G+1}$) so it becomes a trial vector ($u_{i,G+1}$). This vector has a chance to be accepted as the new target with probability Crossover $Cr$ whose binomial uniform formula as follow with the probability of crossover ($Cr$) is chosen from the range between 0 and 1.

$$u_{i,G+1} = \begin{cases} v_{i,G+1}^j & ; \text{rand} < Cr \text{ or } j_{rand} = j \\ x_{i,G}^j & ; \text{other} \end{cases} \tag{4}$$

Afterwards, the fitness function value of the trial vector compares to the value of target vector. If fitness function value of a trial vector is lower than a target vector, then a trial vector is added as a new generation of the population, and vice versa. In this paper, the iteration computation will be terminated when maximum iteration is reached. These criteria make the results as the most optimal for each chosen data. Concretely, in this paper the Absolut Error (AE) is used between the numerical solution $x_{\text{app}}$ and the market price x:

$$\text{AE} = |x_{app} - x| \quad (5)$$

This error measures distances between the computed solution (Black-Scholes result or CmDE result) and the market price one.

**4. Vanilla Option Pricing as a Bi-objective Optimization Problem**

In this section, a vanilla option pricing especially for vanilla call options are modelled as a bi-objective optimization problem and describe the solution methodology that we have used to find its solution.

*4.1. CmDE Algorithm for Calculated an Option Contract*

A two-dimensional space is described and is represented as the search space. Then, there are two parameters in the search space solutions: time of exercising the contract and price of an underlying asset. Because of volatile of financial market and complexity of an option contract, the best solution for this option pricing is not easy to find. Therefore, this behaviour can be captured appropriately using maximum expected option value. Then, we model this problem become bi-objective optimization problem. The maximum expected option value (maximum profit) computes using max ($S - K$, 0) where $S$ is the stock price and $K$ is the strike price, and probability of the options contract calculates using Monte-Carlo simulation.

Pricing a vanilla call option can be more accurate by optimizing the maximum profit and chance of getting that profit. A fitness function value that we use in this paper represents a combination of two objectives that we want to minimize: profit and probability of getting that profit:

$$\text{FitnessValue } (F(x)) = F \text{ (Profit, Probability)} \quad (6)$$

We propose to get the profit and probability of getting the profit simultaneously, in order to for getting the accurate option value. This is called as a bi-objective optimization problem because they are conflicting behaviour from that two objective. The algorithm for this bi-objective optimization problems describes in Algorithm 2.

*4.2. Probability of Computation*

The accurate information can get from the probability of stock price reaching a temporary value in the period of the contract. Because an asset follows a random walk based on efficient market hypothesis (Fama, 1965), it makes not easy to predict the stock price. Therefore, the simulation of Monte-Carlo uses to compute the probability of that stock for getting a temporary value in the future. The algorithm states in (Singh, Thulasiram, & Thulasiraman, 2016).

*4.3 Adaptive weighted-sum method*

Adaptive weighted sum method is one way for optimizing bi-objective problems (Kim, & de-Weck, 2005). According to solutions, feasible search space is evolved in adaptive weighted-sum method. Then, dissimilarity constraints are put on the feasible search space to diminish it by a factor δ for each iteration. δ stands for the space between two extreme points on the proper search space. Therefore, in finding optimal solution, the new region is performed as a sub-optimization. We are doing this procedure until stopping criteria is reached. Finally, Pareto optimal solution is gotten as solution of this multi-objective optimization problem. The algorithm likes in (Singh, Thulasiram, & Thulasiraman, 2016).

**Algorithm 2**. CmDE to assess Vanilla Call Options

| | |
|---|---|
| 1 | CmDE Input parameter $F$ ($mutation\ scale\ factor$), $Cr$ (crossover probability), Population Size ($n$), number of iterations ($t$) |
| 2 | Option Input Parameters: Initial Asset Value ($S$), Expiration Time ($T$), Strike Price ($K$), Volatility ($\sigma$), Risk free rate of interest ($r$) |
| 3 | Randomly initialize position for all vectors $x_i$ where $i = 1,2,...,n$ |
| 4 | For each vector $x_i$ evaluate fitness value $f(x_i)$ using following steps |
| | 4.1 Calculate Pay-off for each vector $x_i$ using equation max(S-K,0) where S represents asset value for each vector |
| | 4.2 For each vector $x_i$ Call procedure PROBCAL such that $Probabilty(x_i) = PROBCAL(S, AssetValue(x_i), Time(x_i), \sigma, r)$ |
| | 4.3 Initialize weight $w_1$ for Pay-off and $w_2$ for Probability |
| | 4.4. Calculate Fitness $F(x_i) = w_1 * Payoff(x_i) + (w_2 * Probability(x_i))$ |
| 5 | The optimum value is represented by fitness value $F(x_i)$ |
| 6 | Call Algorithm 1 to evaluate $F(x_i)$ |
| 7 | Output Global optimum solution |

*4.4. Adaptive weighted-sum method*

We described about using CmDE algorithm with adaptive weighted sum method like in Algorithm 2 for solving option pricing that have changed to a bi-objective optimization problem.

## 5. Experimental Results

We describe the results based on differential evolution algorithm analysis about performance and efficiency.

*5.1. Experimental Setup*

We collect S&P 500 index data for European call option (Historical option data, 2020) which assess in May 31st, 2020 and February 5th, 2021 for data in COVID-19. Then, we compare the experimental results with the BSM results. Then, we assess our differential evolution algorithm about efficiency. We also use the Fed (Historical Federal Interest Rates, 2020) interest rate and CBOE volatility index (VIX) (CBOE, 2020). Besides that, we analyse the data when the covid-19 happened by using implied volatility data that we get from historical option data (Wilmott, 2007). Then, all kinds of contract conditions were introduced, for example, different periods of contract to demonstrate the efficiency of our model in handling all kind of contracts. This is represented in Table 1 and Table 2. Then, a bunch of parametric values uses in our experiments and Table 5 catch on these parametric values.

The second one, we collect data for Netflix that is used for American call option (Wilmott, 2007) which access in June 15th, 2020 and February 5th, 2021 for COVID-19 data. We also used the interest rate from the Fed (Historical Federal Interest Rates, 2020), and implied volatility from Netflix data (Wilmott, 2007). Then, all kinds of contract conditions were introduced, for example, distinct periods of contract to demonstrate the efficiency of our model in dealing with all kinds of contracts. This is represented in Table 3 and Table 4.

**Table 1**. Variation of contracts SPX161216C01925000 before COVID-19

| Expiration Date | Initial Date | $S_0$ | K | The sums of market days | Option Pricing Value in Real Market |
|---|---|---|---|---|---|
| 16 December 2016 | 2 October 2015 | 1948.51 | 1925 | 302 | 143.5 |
| 16 December 2016 | 5 October 2015 | 1987.89 | 1925 | 301 | 169 |
| 16 December 2016 | 6 October 2015 | 1981.01 | 1925 | 300 | 169 |
| 16 December 2016 | 7 October 2015 | 1991.76 | 1925 | 299 | 169 |
| 16 December 2016 | 8 October 2015 | 2012.74 | 1925 | 298 | 169 |

**Table 2**. Variation of contracts SPX200918C02200000 when COVID-19 happen

| Expiration Date | Initial Date | $S_0$ | K | The sums of market days | Option Pricing Value in Real Market |
|---|---|---|---|---|---|
| 18 September 2020 | 1 August 2019 | 2953,56 | 2200 | 297 | 836 |
| 18 September 2020 | 2 August 2019 | 2932.05 | 2200 | 296 | 836 |
| 18 September 2020 | 5 August 2019 | 2844.74 | 2200 | 295 | 836 |
| 18 September 2020 | 6 August 2019 | 2881.77 | 2200 | 294 | 836 |
| 18 September 2020 | 7 August 2019 | 2883.98 | 2200 | 293 | 836 |
| 18 September 2020 | 8 August 2019 | 2938.09 | 2200 | 292 | 836 |

**Table 3**. Variation of contracts NFLX190621C00210000 before COVID-19

| Expiration Date | Initial Date | $S_0$ | K | The sums of market days | Option Pricing Value in Real Market |
|---|---|---|---|---|---|
| 21 June 2019 | 4 January 2018 | 205.63 | 210 | 382 | 36 |
| 21 June 2019 | 5 January 2018 | 209.99 | 210 | 381 | 38.8 |
| 21 June 2019 | 8 January 2018 | 212.05 | 210 | 380 | 40.14 |
| 21 June 2019 | 9 January 2018 | 209.31 | 210 | 379 | 40.14 |
| 21 June 2019 | 10 January 2018 | 212.52 | 210 | 378 | 40 |
| 21 June 2019 | 11 January 2018 | 217.24 | 210 | 377 | 42.98 |

**Table 4**. Variation of contracts NFLX210115C00150000 when COVID-19 happen

| Expiration Date | Initial Date | $S_0$ | K | The sums of market days | Option Pricing Value in Real Market |
|---|---|---|---|---|---|
| 15 January 2021 | 1 August 2019 | 319.5 | 150 | 382 | 186 |
| 15 January 2021 | 2 August 2019 | 318.83 | 150 | 381 | 186 |
| 15 January 2021 | 5 August 2019 | 307.63 | 150 | 380 | 186 |
| 15 January 2021 | 6 August 2019 | 310.1 | 150 | 379 | 165 |
| 15 January 2021 | 7 August 2019 | 304.29 | 150 | 378 | 157 |
| 15 January 2021 | 8 August 2019 | 315.9 | 150 | 377 | 157 |

Table 5. Parameter setting for differential evolution algorithm

| Parameter | Value |
|---|---|
| Population size | 200 |
| Number of iterations | 600 |
| Mutation Scale Factor | 0.5 |
| Crossover Probability | 0.9 |

Our experiments design to represent the (i) usefulness of our CmDE algorithm, (ii) usefulness of our model. Based on Table 1 and Table 2, then Table 3 and Table 4, we can see that the differentiation is in option pricing value in real market data with the time expiration when COVID-19 is higher than before COVID-19 happen.

*5.2. Vanilla Option Contract Results as a Pareto Front*

We compute Pareto front using parametric values provided in Table 5. Based on our experiment, we able to find a set of Pareto optimal solutions. The results show in Figure 1 until Figure 2 for European call option especially for SPX161216C01925000 before COVID-19 and SPX200918C02200000 when COVID-19 happen. Then, we also present the experimental result from Netflix data NFLX190621C00210000 before COVID-19 and NFLX210115C00150000 with time expiration in COVID-19 for American call options. The results present in Figure 5 until Figure 6 before COVID-19 and Figure 7 until Figure 8 when COVID-19 happen.

In Figure 1, we provide solutions for call option contract from 5 October 2015. It is simply noticeable that our differential evolution algorithm effectively finds payoff and probability to achieve that payoff simultaneously. Also, other non-dominant solutions also capture and plot in this figure. Then, based on our experiments on other option contracts show that they behave likewise as presented in other figures. This is an obvious validation of our differential evolution algorithm in calculating an option contract when formulated as a bi-objective optimization problem. Since noting the real pay-off received for each contract, we calculate the absolute error for all the solutions. The results show in Table 6 until Table 7 for S&P 500 Index and Table 8 until Table 9 for Netflix.

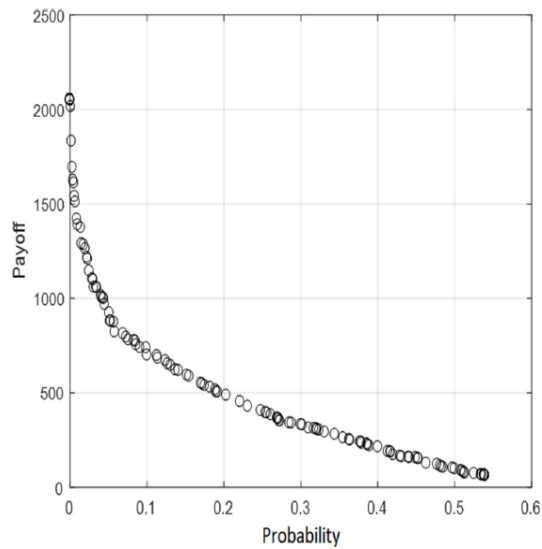

**Figure 1.** Pareto front for SPX161216C01925000 on October 5$^{th}$, 2015 before COVID-19.

From Figure 1, we get that the option value is 168.90 (with the initial conditions of $S_0 = 1987.89$, and $K = 1925$). The market option value for the contract was found to be 169. Therefore, the % error is 0.06%. Then, we show the comparison results in finding the option value by CmDE, DE, and Black-Scholes (BS) as shown in Table 6. The result of the BS for the same contract is 205.68, then the result of DE is 171.4, and the result of CmDE is 168.90. The percentage of error by the BS, DE, and CmDE are 21.70%, 1.42%, 0.06% . We get the CmDE is the best algorithm in finding the option value which nearest to the option pricing in the market because it has the smallest error. Then, our CmDE able to find the true option values in the Pareto front like in Figure 1 and Figure 2 for European call option pricing before COVID-19. This shows that our CmDE algorithm with adaptive weighted sum method of option pricing using probability and pay-off as objectives, are very efficient and able to find on the accurate real market scenario. Pareto optimal solutions found using our strategy catch on all the possible payoff that we get from contract. This approves the efficiency of our model.

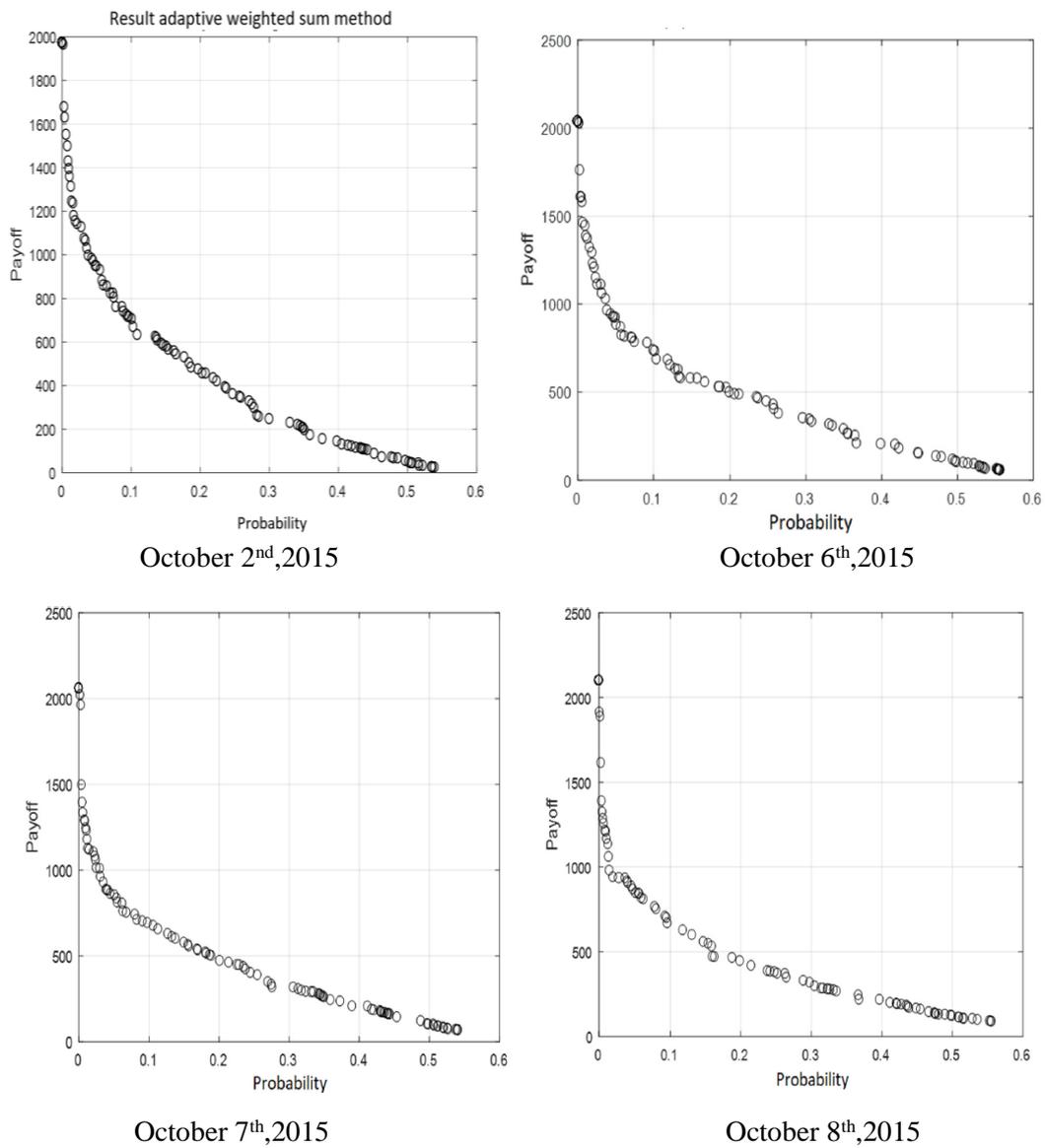

October 2$^{nd}$,2015      October 6$^{th}$,2015

October 7$^{th}$,2015      October 8$^{th}$,2015

**Figure 2.** Pareto front for SPX161216C01925000 on October 2$^{nd}$, 2015 until October 8$^{th}$, 2015 before COVID-19.

**Table 6.** Comparison the Experimental Result of SPX161216C01925000 using CmDE Algorithm, DE Algorithm and B-S before COVID-19

| Date | Option Pricing in Market | B-S Results | DE Results | CmDE Results | %Absolute Error of B-S | %Absolute Error of DE | %Absolute Error of CmDE |
|---|---|---|---|---|---|---|---|
| 2 October 2015 | 143.5 | 194.26 | 143.5 | 143.20 | 35.37 | 0.00 | 0.21 |
| 5 October 2015 | 169 | 205.68 | 171.4 | 168.90 | 21.70 | 1.42 | 0.06 |
| 6 October 2015 | 169 | 200.03 | 153.4 | 166.50 | 18.36 | 9.23 | 1.48 |
| 7 October 2015 | 169 | 198.03 | 169.8 | 168.00 | 17.18 | 0.47 | 0.59 |
| 8 October 2015 | 169 | 202.90 | 167.6 | 168.90 | 20.06 | 0.83 | 0.06 |

**Table 7.** Comparison the Experimental Result of SPX200918C02200000 with COVID-19 data

| Date | Option Pricing in Market | B-S Results | DE Results | CmDE Results | %Absolute Error of B-S | %Absolute Error of DE | %Absolute Error of CmDE |
|---|---|---|---|---|---|---|---|
| 1 August 2019 | 836 | 818.41 | 829.06 | 834.8 | 2.10 | 0.83 | 0.14 |
| 2 August 2019 | 836 | 786.73 | 837.95 | 837.7 | 5.89 | 0.23 | 0.20 |
| 5 August 2019 | 836 | 700.27 | 834.82 | 832.7 | 16.24 | 0.14 | 0.39 |
| 6 August 2019 | 836 | 781.08 | 823.29 | 834.9 | 6.57 | 1.52 | 0.13 |
| 7 August 2019 | 836 | 738.32 | 836.19 | 829.5 | 11.68 | 0.02 | 0.78 |
| 8 August 2019 | 836 | 774.54 | 830.22 | 833 | 7.35 | 0.69 | 0.36 |

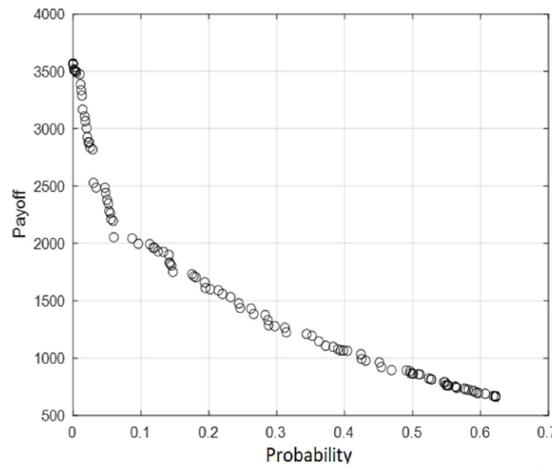

**Figure 3.** Pareto front for SPX200918C02200000 on August 6[th], 2019 with COVID-19 data.

We continue about the result of our algorithm in finding option pricing of SPX 200918C02200000 with data when COVID-19 happen. The results show in Figure 3 and Figure 4. From Figure 3, we see that the option value is 834.90 (with the initial conditions of $S_0 = 2881.77$ and $K = 2200$). The actual option value for the same contract was found to be 836. Therefore, the % error is 0.13%. Then, we show the comparison results in finding the option value by CmDE, DE, and Black-Scholes (BS) as shown in Table 7. The result of the BS for the same contract is 781.08, then the result of DE is 823.29, and the result of CmDE is 834.9. The percentage of error by the BS, DE, and CmDE are 6.57%, 1.52%, 0.13% . We get the CmDE is the best algorithm in finding the option value which nearest to the option pricing in the market because it has the smallest error. Our

CmDE algorithm able to find the true option values in the Pareto front like in Figure 3 and Figure 4 for European call option pricing with COVID-19 data. This shows that our CmDE algorithm with adaptive weighted sum method very efficient and able to find the accurate real market scenario. Then, our strategy can find all Pareto optimal solutions to show payoff and probability for achieving that payoff. This is validating the efficiency of our model.

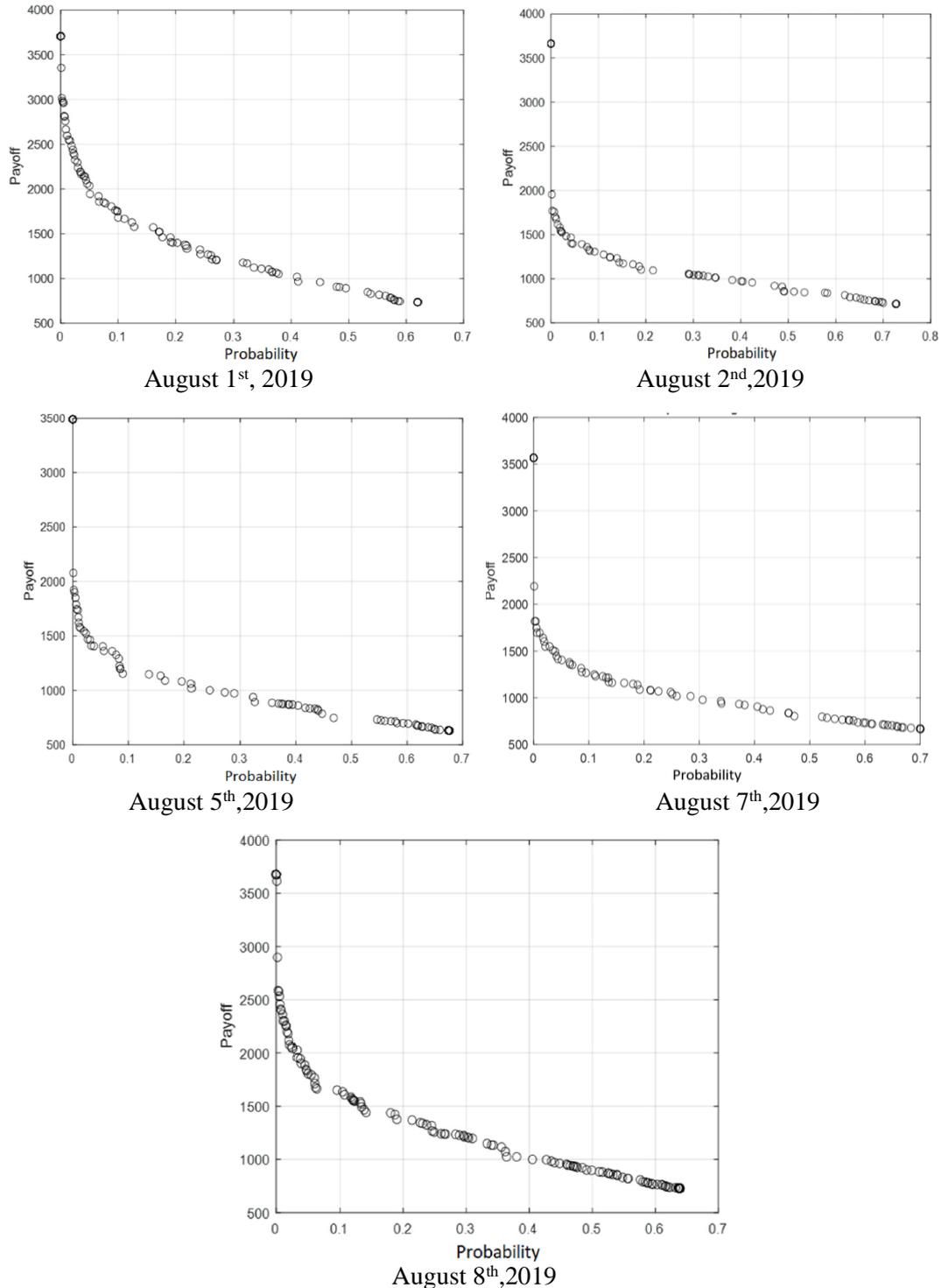

**Figure 4.** Pareto front for SPX200918C02200000 on August 1st, 2019 until August 8th, 2019 with COVID-19 data.

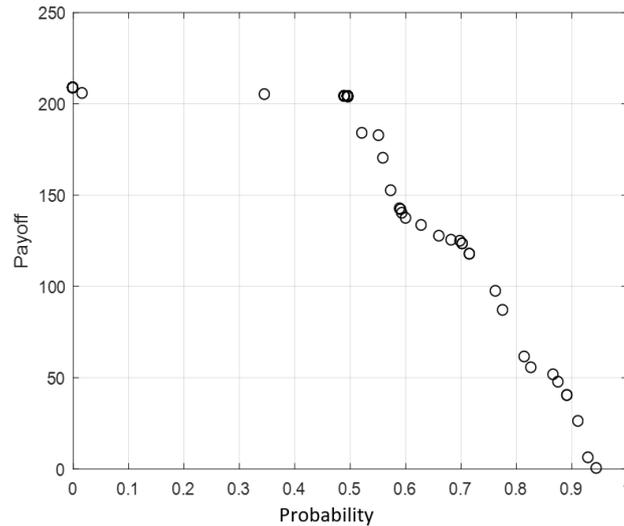

**Figure 5.** Pareto front for NFLX190621C00210000 on January 9th, 2018 before COVID-19.

We continue about the result of our algorithm in finding option pricing of Netflix NFLX190621C00210000 before COVID-19. The results show in Figure 5 and Figure 6. From Figure 5, we see that the option value is 40,99 (with the initial conditions of $S_0$ = 209.31 and $K$ = 210). The actual option value for the same contract is 40.14. Therefore, the % error is 2.12%. Then, we show the comparison results in finding the option value by CmDE, DE, and Black-Scholes (BS) as shown in Table 8. The result of the BS for the same contract is 38.59, then the result of DE is 41.57, and the result of CmDE is 40.99. The percentage of error by the BS, DE, and CmDE are 3.86%, 3.56%, 2.12%. We get the CmDE is the best algorithm in finding the option value which nearest to the option pricing in the market because it has the smallest error. Then, our CmDE algorithm able to represent the true option values in the Pareto front like in Figure 6 for NFLX190621C00210000 as American call option pricing before COVID-19. This shows that our CmDE algorithm with adaptive weighted sum method of option pricing using probability and payoff as objectives, are very efficient and able to represent the accurate real market scenario.

We continue about the result of our algorithm in finding option pricing of NFLX210115C00150000 with COVID-19 data. The results are showed in Figure 7 and Figure 8.

From Figure 7, we see that the option value is 186.07 (with the initial conditions of $S_0$=318.83 and K=150). The actual option value for the same contract was found to be 186. Therefore, the % error is 0.04%. Then, we show the comparison results in finding the option value by CmDE, DE, and Black-Scholes (BS) as shown in Table 9. The result of the BS for the same contract is 178.44, then the result of DE is 184.02, and the result of CmDE is 186.07. The percentage of error by the BS, DE, and CmDE are 4.06%, 1.06%, 0.04%. We get the CmDE is the best algorithm in finding the option value which nearest to the option pricing in the market because it has the smallest error. Then, our CmDE algorithm able to represent the true option values in the Pareto front like in Figure 7 and 8 for Netflix NFLX210115C00150000 as American call option pricing without paying dividend with COVID-19 data. This shows that our CmDE algorithm with adaptive weighted sum method of option pricing using probability and payoff as objectives, are very efficient and able to represent the accurate real market scenario. Our strategy can represent all the pareto front from the contract. This is validating the efficiency of our model.

The experimental result using CmDE algorithm compare with the option pricing in market data, Black-Scholes (BS) formula, and Differential Evolution (DE) algorithm. The comparison shows in Table 6 until Table 7 for European Call option before CoVID-19 and after COVID-19 happen. Then, the comparison results for American call option without paying dividend shows in Table 8 until Table 9 for data before COVID-19 and after COVID-19 happen.

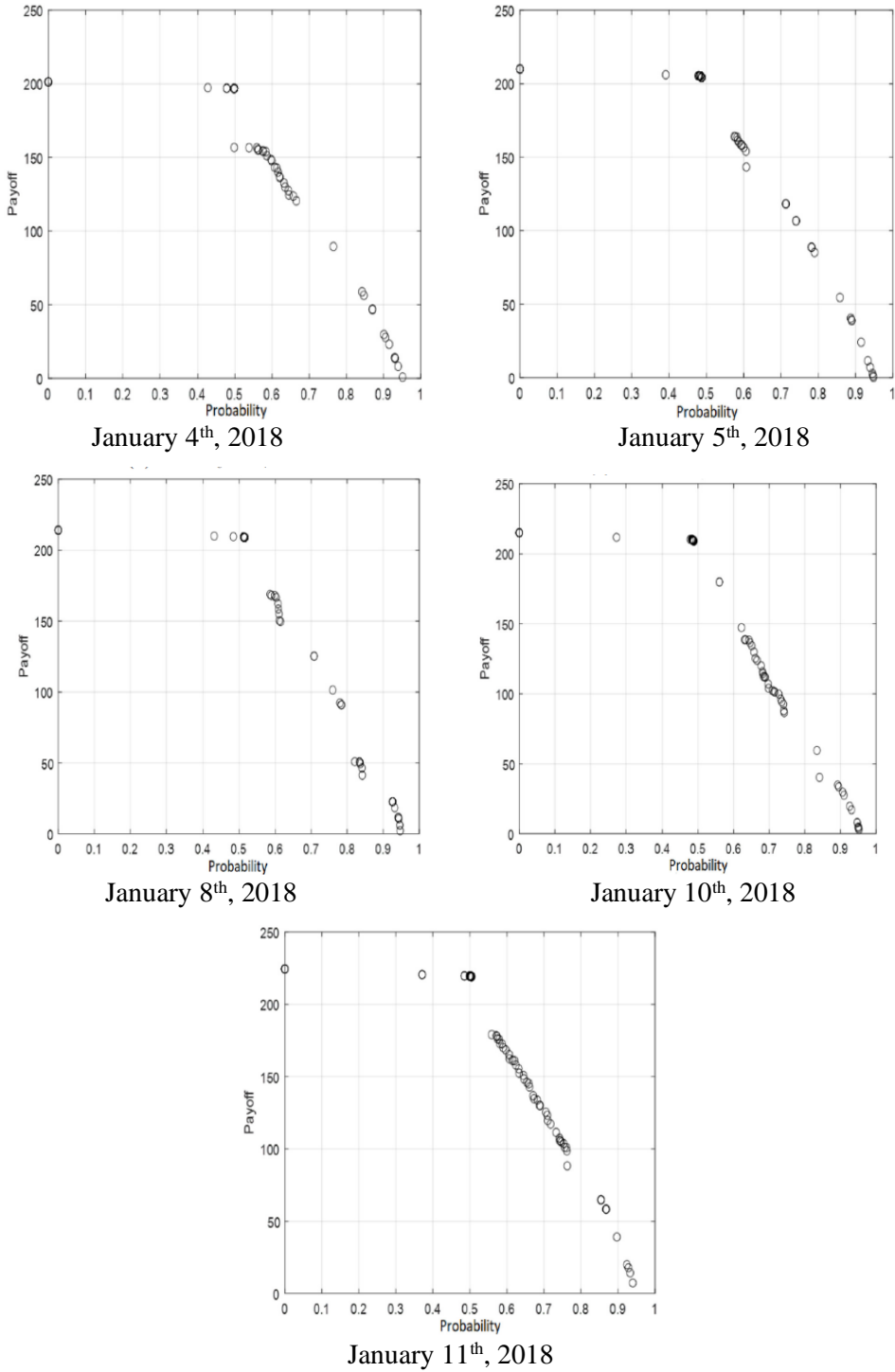

January 4th, 2018

January 5th, 2018

January 8th, 2018

January 10th, 2018

January 11th, 2018

**Figure 6.** Pareto front for NFLX190621C00210000 on January 4th, 2018 until January 11th, 2018 before COVID-19.

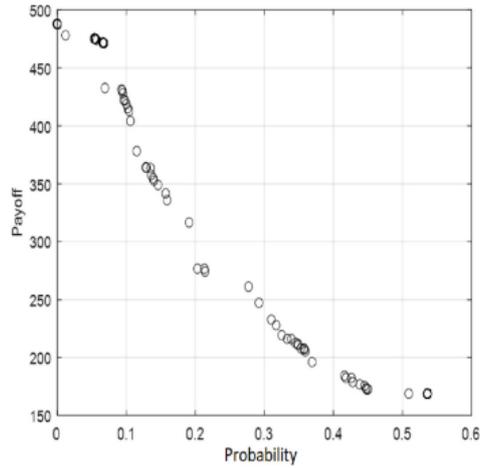

**Figure 7.** Pareto front for NFLX210115C00150000 on August 2$^{nd}$,2019 with COVID-19 data.

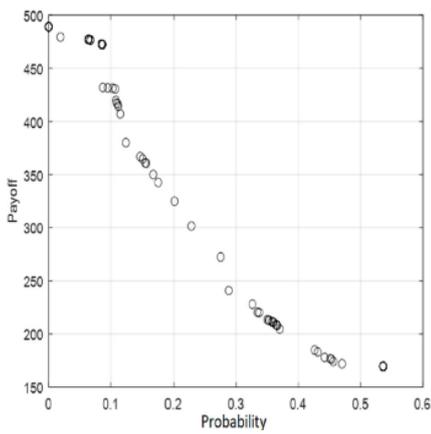
August 1$^{st}$, 2019

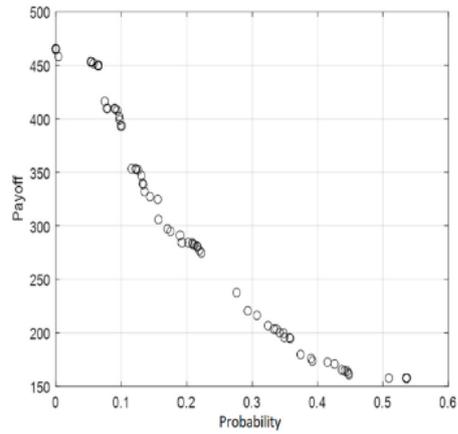
August 5$^{th}$,2019

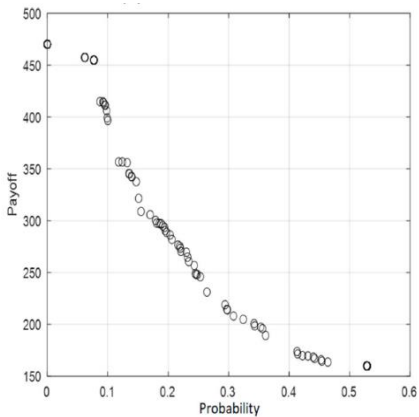
August 6$^{th}$,2019

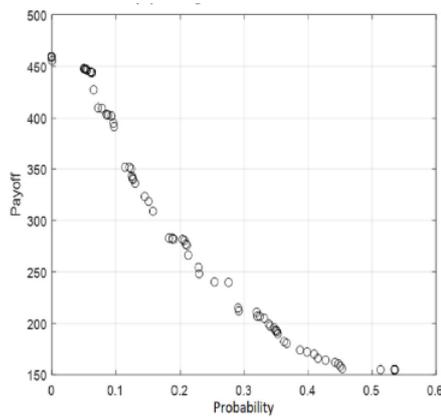
August 7$^{th}$,2019

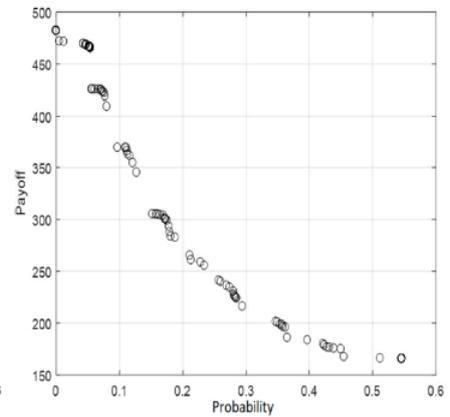
August 8$^{th}$,2019

**Figure 8.** Pareto front for NFLX210115C00150000 on August 2$^{nd}$, 2019 until August 8$^{th}$, 2019 with COVID-19 data

Based on data in Table 6 and Table 7, the option pricing value for European call option for S&P 500 Index in the market after COVID-19 happen is higher than the option pricing value before COVID-19 happen. This is because the stock price value after COVID-19 happen is higher than the stock price value before COVID-19 happen. It can be said that the price of the option is directly proportional to the price of the stock. Therefore, when the stock price goes up, the price of stock options will also go up. We also can see this situation for European call option using S&P 500 Index and American call option without paying dividend using Netflix stock.

Based on the data in Table 8 and Table 9, we can see that the Netflix's stock price was lower before COVID-19 occurred compared to Netflix's stock price at the time of COVID-19. This happened because during COVID-19 many people did not work in offices, school children did not come to school, college students did not come to college and almost everyone doing activity from home. Therefore, Netflix, as a subscription-based streaming service that allows Netflix members to watch TV shows and movies on devices connected to the Internet, get the stock price value become higher than before COVID-19 happen. Many people need entertainment when doing activity at home. Therefore, they choose Netflix as tools to reduce stress while working from home. This is what made Netflix's stock price go up during the COVID-19. We conclude that COVID-19 has made the price of European call options of the S&P 500 Index and the price of American call option from Netflix become higher than before.

**Table 8.** Comparison the Experimental Result of NFLX190621C00210000 using CmDE Algorithm, DE Algorithm and B-S before COVID-19

| Date | Option Pricing in Market | B-S Results | DE Results | CmDE Results | %Absolute Error of B-S | %Absolute Error of DE | %Absolute Error of CmDE |
|---|---|---|---|---|---|---|---|
| 4 January 2018 | 36 | 36.26 | 35.54 | 35.54 | 0.72 | 1.28 | 1.28 |
| 5 January 2018 | 38.8 | 39.00 | 38.41 | 38.41 | 0.52 | 1.01 | 1.01 |
| 8 January 2018 | 40.14 | 40.85 | 40.56 | 40.56 | 1.77 | 1.05 | 1.05 |
| 9 January 2018 | 40.14 | 38.59 | 41.57 | 40.99 | 3.86 | 3.56 | 2.12 |
| 10 January 2018 | 40 | 40.27 | 39.72 | 40.58 | 0.68 | 0.70 | 1.45 |
| 11 January 2018 | 42.98 | 43.66 | 39.57 | 41.03 | 1.58 | 7.93 | 4.54 |

**Table 9.** Comparison the Experimental Result of NFLX210115C00150000 with COVID-19 data

| Date | Option Pricing in Market | B-S Results | DE Results | CmDE Results | %Absolute Error of B-S | %Absolute Error of DE | %Absolute Error of CmDE |
|---|---|---|---|---|---|---|---|
| 1 August 2019 | 186 | 178.47 | 187.53 | 187.93 | 4.05 | 0.82 | 1.04 |
| 2 August 2019 | 186 | 178.44 | 184.02 | 186.07 | 4.06 | 1.06 | 0.04 |
| 5 August 2019 | 186 | 166.11 | 185.14 | 186.53 | 10.69 | 0.46 | 0.28 |
| 6 August 2019 | 165 | 168.83 | 163.14 | 165.71 | 2.32 | 1.13 | 0.43 |
| 7 August 2019 | 157 | 162.78 | 155.80 | 156.36 | 3.68 | 0.76 | 0.41 |
| 8 August 2019 | 157 | 173.12 | 160.31 | 160.19 | 10.27 | 2.11 | 2.03 |

## 6. Conclusions

Combinational Mutation Strategy of Differential Evolution (CmDE) algorithm with adaptive weighted sum method use in finding Pareto optimal solution of vanilla option pricing that has been modelled into a bi-objective optimization problem. We focus on bi-objective optimization problem because the objective functions are payoff and probability to achieve that payoff in vanilla option pricing problems. Then, an investor can use the result of our model as consideration.

Based on our experiments, we can show the competency of our model and accuracy and efficiency of our algorithm to price vanilla option. The results compute based on these experiments can capture the real market value appropriately. Then, every Pareto front that we get capturing the true option value in the Pareto front. Therefore, our algorithm competent to find a solution that is common to option price from the real market data. The results also show that our metaheuristics algorithm can use to predict the option with the time expiration in COVID-19 happen. Thus, an investor can use our model and algorithm as strategy to assess the option contract before dealing with the contract.